# Deep Hierarchical Classification for Category Prediction in E-commerce System

Dehong Gao, Wenjing Yang, Huiling Zhou, Yi Wei, Yi Hu and Hao Wang
Alibaba Group
Hangzhou City, Zhejiang Province, China
{dehong.gdh, carrie.ywj, zhule.zhl, yi.weiy, erwin.huy, longran.wh}@alibaba-inc.com

## Abstract

In e-commerce system, category prediction is to automatically predict categories of given texts. Different from traditional classification where there are no relations between classes, category prediction is reckoned as a standard hierarchical classification problem since categories are usually organized as a hierarchical tree. In this paper, we address hierarchical category prediction. We propose a Deep Hierarchical Classification framework, which incorporates the multi-scale hierarchical information in neural networks and introduces a representation sharing strategy according to the category tree. We also define a novel combined loss function to punish hierarchical prediction losses. The evaluation shows that the proposed approach outperforms existing approaches in accuracy.

## 1 Introduction

Category Prediction (CP), which aims to recognize the intent categories of given texts, is regarded as one of the most fundamental machine learning tasks in e-commerce system (Ali et al., 2016). For example, this predicted category information will influence product ranking in search and recommendation system.

Different from the traditional classification (Yann et al., 1998; Larkey and Croft, 1996) CP is formally categorized as a hierarchical classification task since categories in most e-commerce websites are organized as a hierarchical tree (we consider the situation that the categories are organized as a hierarchical tree, but not a directed acyclic graph). Figure 1.(a) shows a simplified fragment of one category architecture. Apart from CP, there are also many other tasks belonging to hierarchical classification, e.g., image classification shown in Figure 1.(b).

For simplicity, most practical approaches ignore the relation information between classes (hereafter referred to as flat classification). These approaches are easily implemented, but disadvantage in accuracy (Rohit et al., 2013). In academy, the hierarchical classification problem is not well-studied as well (Silla and Freitas., 2011). Except these flat approaches, published studies are mainly divided into two directions: the local approaches, and the global approaches (Carlos and Freitas., 2009). The local approaches learn multiple independent classifiers, each classifier either for per node, or for per parent node or for per layer. Taking the local approach for per layer as an example, for Figure 1.(b) it will train two independent classifiers for layer_1 and layer_leaf, respectively. The global approaches regard all none-root nodes as the classes to predict. Only one classifier is trained for all these none-root classes. We argue that all these approaches either do not consider the hierarchical structure at all (i.e., the flat approaches), or take implicit or tiny consideration of the class hierarchy.

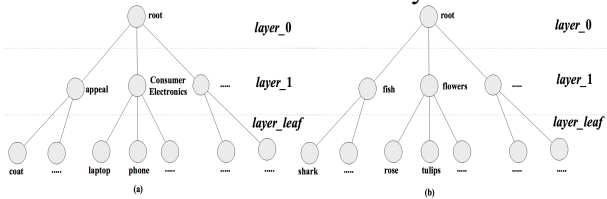

Figure 1. Hierarchical Classification Tasks

The main challenges in hierarchical classification are at two aspects: hierarchical representation in classification model and hierarchical inconsistency in training process. Hierarchical representation means researchers may select Naive Bayesian (Larkey and Croft, 1996), Support Vector Machine (Chang and Lin, 2009), and Neural Networks (Jurgen Schmidhuber, 2015) as their classification models. But the hierarchical information fails to be explicitly incorporated in these models. Consequentially, it is hard for these models to learn the complex hierarchical information. The hierarchical inconsistency means if a text is predicted as "appeal" in the layer_1, but as "laptop" in the leaf layer during training phase in Figure 1.(a), none approach can deal with this inconsistency as far as we known.

To solve these two problems, we propose a general Deep Hierarchical Classification (DHC) framework. Firstly, according to hierarchical representation, our DHC approach directly incorporates class hierarchy information in neural network. DHC first generates one hierarchical layer representation for per layer. Inspired by the idea that sibling classes under one parent class must share certain common information, we introduce a hierarchical representation sharing strategy that the representation of one lower layer should include the representation information about its upper layer. This sharing strategy is recursively carried on in a top-down manner according to the class hierarchy. As a result, the classification model is forced to learn this structure information, and the class hierarchy information is "explicitly" involved in the model. Secondly, according to hierarchical inconsistency, we define a hierarchical loss function composed of the layer loss and the dependence loss. The layer loss defines the training loss within layers, which is the same to the loss in traditional flat classification. The proposed dependence loss defines the loss between layers. When predictions of two successive layers are inconsistent (i.e., these two predicted classes are not in a parent-child relationship), we will add an additional dependence loss to compel the classification model to learn this relation information. The dependence loss function is hierarchy-related and is regarded as a punishment when predictions are not consistent with the category structure. By this way, we can deal with the hierarchical inconsistency during the training process.

DHC can be regarded as a general hierarchical classification framework, we evaluate it with text and image classification. For text classification, we collect query-category and title-category pairs from one e-commence website. For image classification we adopt the commonly-used cifar100 dataset. Taking advantage of hierarchical representation and hierarchical loss function, the DHC approach significantly improves the accuracy. Our main contributions include the novel DHC framework and the hierarchical representation and hierarchical loss which are first proposed as far as we know. All of them will be detailed in the following sections.

## 2 Deep Hierarchical Classification

Mathematically, the hierarchical classification task can be formulated as: <u>Given</u>:
**Input $X$**: $X$ can denote the text or the image inputs.
**Category tree $\mathcal{T}$**: Categories are organized by a category tree $\mathcal{T}$ with $L$ hierarchical layers. The categories (i.e., classes to predict) are denoted by $Y$. Categories of different layers are dependent as $Y_L \Rightarrow Y_{L-1} \Rightarrow \cdots \Rightarrow Y_1$ ( $\Rightarrow$ denotes the IS-A relation in category tree $\mathcal{T}$.)
<u>Output</u>: **Categories of input $X$**: Predict categories of the given input $X$. Since categories are hierarchically related, it is possible to predict the leaf class and infer the classes of all the other layers according to category tree $\mathcal{T}$.

In the DHC approach, we defines a neural network model $\mathcal{N}(\theta)$ where $\theta$ are the parameters to be estimated. Taking a three-layer hierarchical classification problem as an example, we show the DHC neural network in Figure 2. The neural network is composed of three parts: **Flat Neural Network (FNN)**, **Hierarchical Embedding Network (HEN)** and **Hierarchical Loss Network (HLN)**. We will further discuss these three parts.

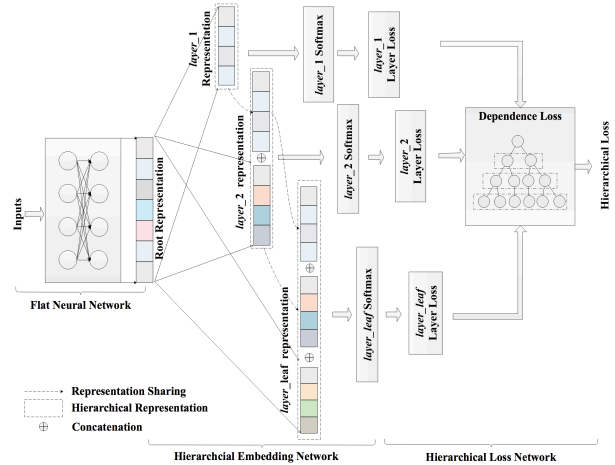

Figure 2. Deep Hierarchical Classification (Take three-layer hierarchical classification as example)

### 2.1 Flat Neural Network

Given an input $X$, FNN is used to generate a root representation. For our main contributions lay in the HEN and HLN, we can adopt a state-of-the-art neural network in practice. Let $\mathcal{N}_{flat}(\theta_{flat})$ denote this flat neural network, the output is viewed as the root representation $R_0$

$$R_0 = \mathcal{N}_{flat}(X, \theta_{flat}) \qquad (1)$$

## 2.2 Hierarchical Embedding Network

With the root representation, HEN aims to produce hierarchical representations for every layers. For the $l^{th}$ layer, we first produce the independent representation $R'_l$, i.e.,

$$R'_l = W_{r_l} * R_0 \quad (2)$$

where $W_{r_l}$ represents the weights for the independent layer representation. The independent layer representation is hierarchical-free. As mentioned, classes belonging to the same parent class share certain common information. The representation of one lower layer should include the representation information about its upper layer. Thus, the hierarchical representation $R_l$ is computed by concatenating the independent representations of all previous layers denoting by

$$\begin{cases} R_l = R_{l-1} \oplus R'_l & for\ l \neq 1 \\ R_l = R'_l & for\ l = 1 \end{cases} \quad (3)$$

For the $l^{th}$ layer prediction, the hierarchical representation $R_l$ is passed into a softmax regression layer. The output of the softmax regression layer is denoted by

$$\tilde{y}_{li} = \frac{e^{W_{sli}R_l}}{\sum_{k=1}^{|l|} e^{W_{slk}R_l}} \quad (4)$$

where $W_{s_l}$ are the parameters of the $l^{th}$ softmax regression layer. $\tilde{y}_{li}$ denotes the prediction probability of the $i^{th}$ class in the $l^{th}$ hierarchy layer. $|l|$ denotes the number of classes in the $l^{th}$ hierarchy layer.

## 2.3 Hierarchical Loss Network

According to hierarchical layer representations and document true classes, HLN will compute the hierarchical loss to estimate the neural network parameters. We propose two types of losses, i.e., the layer loss and the dependency loss. Concretely, the $l^{th}$ layer loss function $lloss_l$ is defined as

$$lloss_l = -\sum_{j=0}^{|l|} y_{lj} \log(\tilde{y}_{lj}) \quad (5)$$

$y_{lj}$ is the expected output of the $j^{th}$ class in the $l^{th}$ hierarchy layer. To measure the prediction errors between layers, we propose a dependence loss. If the predicted classes of two successive layers are not parent-child relation, a dependence loss appears to punish the learning model for it does not predict the classes according to the hierarchy structure. The $l^{th}$ layer dependence loss $dloss_l$ is defined as

$$dloss_l = -(ploss_{(l-1)})^{\mathbb{D}_l \mathbb{I}_{l-1}} (ploss_l)^{\mathbb{D}_l \mathbb{I}_l} \quad (6)$$

where $\mathbb{D}_l$ and $\mathbb{I}_l$ denote that whether the model predictions conflict category structure, especially

$$\mathbb{D}_l = \begin{cases} 1 & if\ \hat{y}_l \not\Rightarrow \hat{y}_{l-1} \\ 0 & else \end{cases}$$
$$\mathbb{I}_l = \begin{cases} 1 & if\ \hat{y}_l \neq y_l \\ 0 & else \end{cases} \quad (7)$$

Here $\hat{y}_l = \max_i \tilde{y}_{li}$ denotes the predicted class, and $y_l$ is the true label of the query. $\mathbb{D}_l$ denotes whether the predicted label in the $l^{th}$ layer is a child class of the predicted class in the $l - 1^{th}$ layer. $\mathbb{I}_l$ denotes whether the $l^{th}$ layer prediction is correct. $ploss_l$ is a dependence punishment to force the neural network to learn structure information from the category structure. $ploss_l$ can be set as a constant or related to the prediction error.

Finally, the total loss is defined as the weighted summation of the layer losses and the dependence losses, i.e.,

$$J(\theta) = \sum_{i=1}^{L} \alpha_i lloss_i + \sum_{i=2}^{L} \beta_i dloss_i \quad (8)$$

where $\alpha$ and $\beta$ ($0 \leq \alpha, \beta \leq 1$) are the loss weights of different layers.

In the inference phase, there are mainly three methods to determine the category of one text, the heuristic method, the greedy method and the beam search method (Wu et al., 2016). We adopt the greedy method in our experiments for fair comparison.

| Datasets | Sample | 1st&2nd layer |
|---|---|---|
| Query-Category | 1.3millions | 39/742 |
| Title-Category | 30.7millions | 39/742 |
| Cifar100[1] | 60thousands | 20/100 |

Table 1. Information of experiment datasets

## 3 Experiments

### 3.1 Datasets

As DHC is a general hierarchical classification framework, we experiment on text classification and image classification with both industry and public datasets, respectively. For text classification, we collect two datasets, i.e., <Query-Category> (user query and the category of one user-clicked product) and <Title-Category> (product title and its category). For image classification, we experiment on the cifar100 dataset, in which the fine and coarse labels are organized by a three-layer hierarchical tree. The information of these three datasets (e.g., sample numbers and class numbers) are shown in Table 1.

---
[1] https://www.cs.toronto.edu/~kriz/cifar.html

For text classification, query-category and title-category corpus are randomly divided into ten equal parts. Nine parts are used in the training phase and the other one is used in the test phase. For image classification, we use the official training/testing parts. Accuracy is selected to evaluate the performances (Kiritchenko and Stan, 2005; Kiritchenko and Stan, 2006).

| Accuracy | Query-Category | | Title-Category | |
|---|---|---|---|---|
| | 1st layer | 2nd layer | 1st layer | 2nd layer |
| SVM | 88.1% | 67.99% | 85.34% | 60.13% |
| HSVM | 89.98% | 68.59% | 88.14% | 63.59% |
| FastText | 90.10% | 67.64% | 88.06% | 61.62% |
| TextCNN | 90.11% | 68.29% | 89.10% | 64.31% |
| HiNet | 91.54% | 72.98% | 90.69% | 65.10% |
| DHC | **92.10%** | **73.37%** | **91.21%** | **69.02%** |

Table 2. Accuracy evaluation of SVM, HSVM, FastText, TextCNN, HiNet and DHC approaches for text classification

| Accuracy | Cifar100 | |
|---|---|---|
| | 1st layer | 2nd layer |
| KerasCNN | 89.23% | 67.89% |
| HiNet | 90.11% | 72.23% |
| DHC | **92.21%** | **75.91%** |

Table 3. Accuracy evaluation of baseline, HiNet and DHC approaches for image classification

### 3.2 Evaluation of baseline and existing approaches

In this set of experiments, we compare our approach with the existing approaches.

For text classification, SVM (Chang and Lin, 2009), FastText (Joulin et al., 2016), and TextCNN (Yoon Kim, 2014) are selected as the flat baselines and we train two classifiers for the two layers, respectively. HSVM (Tsochantaridis et al., 2005) and HiNet (Wu and Saito, 2017) are selected as hierarchical baselines. For fair competition, HiNet and DHC are adopted the same network architecture with TextCNN as the base model. The purpose is to verify the effectiveness of our DHC framework, but not the based model.

With the limited space, the standard neural network (KerasCNN)[2] and HiNET are adopted as the flat and hierarchical baselines in image classification, respectively. HiNET and DHC keep the same network architecture and hyper-parameters with KerasCNN. We also focus on the comparison of the DHC framework, rather than the base model.

The accuracies of these four approaches are shown in Table 2 and Table 3, which shows that DHC outperforms all the other approaches. The layer representation sharing and hierarchical loss computation help the improvement in performance. Meanwhile, we find that the accuracy increase of the leaf layer is greater than that of the layer_1. This is because the classification for the layer_1 is much easier than that for the leaf layer. The classifiers can learn comparable models for the layer_1, but DHC shows its powerful ability in the leaf layer classification.

### 3.3 Evaluation of HEN and HLN

This set of experiments is to reveal the influence of HEN and HLN. HiNet is adopted as the baseline approach and the experiments are conducted on title-category dataset for simplicity.

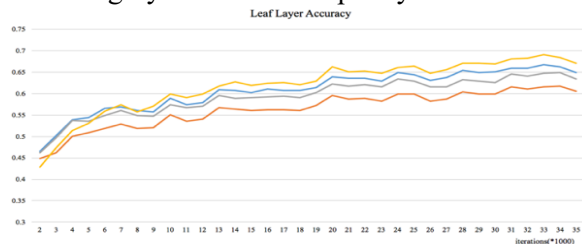

Figure 3. Accuracy evaluation of HiNet, DHC_HEN ($\beta = 0$ in Equation 8), DHC_HLN ($R_l = R'_l$ in Equation 3) and DHC approaches

Figure 3 illustrates the accuracy changes of the leaf layer prediction in the training iteration. Compared to HiNet, it indicates that both HEN and HLN have the positive influence for hierarchical classification. HEN contributes more than HLN. We find that the definition of the hierarchical loss function affects the robustness and accuracy of the classification a lot. A proper hierarchical loss function definition is still an open question.

## 4 Conclusions

In sum, we extensively address the two challenges (i.e., hierarchical representation and hierarchical inconsistency) in hierarchical classification and propose the DHC approach to solve these two problems. Experiments both on text and image classification demonstrate the effectiveness of our proposed DHC approach.

---
2 https://keras.io/examples/cifar10_cnn/